\documentclass{article}
\usepackage{spconf,amsmath,graphicx}
\usepackage{booktabs,arydshln,multirow,multicol,algorithm,algorithmic,url}

\title{UniX-Encoder: A Universal $X$-Channel Speech Encoder for Ad-Hoc Microphone Array Speech Processing}
\name{\begin{tabular}{c}Zili Huang$^{1\dagger}$, 
Yiwen Shao$^1$, 
Shi-Xiong Zhang$^2$,
Dong Yu$^2$
\end{tabular}\thanks{$^\dagger$Work performed during an internship at Tencent AI Lab.}}
\address{   
$^1$Center for Language and Speech Processing, Johns Hopkins University, Baltimore, MD, USA\\
$^2$Tencent AI Lab, Bellevue, WA, USA\\}
\begin{document}
\ninept
\maketitle
\begin{abstract}

The speech field is evolving to solve more challenging scenarios, such as multi-channel recordings with multiple simultaneous talkers. Given the many types of microphone setups out there, we present the UniX-Encoder.  It's a universal encoder designed for multiple tasks, and worked with any microphone array, in both solo and  multi-talker environments. Our research enhances previous multi-channel speech processing efforts in four key areas:
1) Adaptability: Contrasting traditional models constrained to certain microphone array configurations, our encoder is universally compatible. 2) Multi-Task Capability: Beyond the single-task focus of previous systems, UniX-Encoder acts as a robust upstream model, adeptly extracting features for diverse tasks including ASR and speaker recognition. 3) Self-Supervised Training: The encoder is trained without requiring labeled multi-channel data. 4) End-to-End Integration: In contrast to models that first beamform then process single-channels, our encoder offers an end-to-end solution, bypassing explicit beamforming or separation.
To validate its effectiveness, we tested the UniX-Encoder on a synthetic multi-channel dataset from the LibriSpeech corpus. Across tasks like speech recognition and speaker diarization, our encoder consistently outperformed combinations like the WavLM model with the BeamformIt frontend.
\end{abstract}
\vspace{-.5em}
\section{Introduction}
\vspace{-.5em}
\label{sec:intro}
Conversational speech processing presents a formidable challenge, primarily owing to the presence of pervasive noise, reverberation, and overlapping speech from multiple speakers~\cite{ccetin2006analysis,chen2020continuous,raj2021integration}. In recent years, the field has witnessed a surge in innovative deep learning-based approaches, encompassing both single-channel~\cite{kanda2020serialized,kanda2022streaming} and multi-channel~\cite{chang2019mimo,watanabe2020chime,medennikov2020target,zhang2021adl,zhang2022all,yoshioka2022vararray,kanda2023vararray} methods, to effectively tackle a spectrum of issues related to conversational speech, spanning from speech recognition and speech separation to speaker diarization.

In contrast to single-channel-based systems, multi-channel-based speech processing systems can further harness spatial information, resulting in an improved capacity to discriminate between sources. The design of multi-channel speech processing systems exhibits remarkable diversity, encompassing both modular~\cite{watanabe2020chime,medennikov2020target,zhang2021adl,yoshioka2022vararray,kanda2023vararray} and end-to-end~\cite{chang2019mimo,subramanian2021directional} approaches.

Despite significant accomplishments in numerous downstream tasks, current multi-channel speech processing systems exhibit several notable limitations. Firstly, a majority of these systems are reliant on specific devices, rendering them challenging to adapt for use with alternative microphone arrays. Secondly, many existing systems necessitate supplementary information, such as microphone topology and target direction, which may not be readily accessible in practical applications. Lastly, a substantial portion of these systems relies on clean sources or text transcripts as their supervision targets, thereby precluding the utilization of unlabeled web-based multi-channel speech data.

To tackle the previously mentioned challenges, we introduce the UniX-Encoder (UniX-Enc), a self-supervised learning-based model tailored for multi-channel speech processing. Our approach ensures device independence and task agnosticism during the initial pretraining phase, thus enabling its versatile application across a wide range of scenarios. Subsequently, during the fine-tuning stage, we further optimize the representations for specific downstream applications.

Rather than introducing a novel framework aimed solely at addressing the particular challenges of multi-channel speech recognition or speaker diarization, our research centers on the extraction of potent representations from multi-channel speech data. These robust representations are envisioned as a valuable foundation for subsequent models, with the potential to significantly augment system performance in various downstream tasks.

We assess the efficacy of our novel UniX-Enc model using a synthetic multi-channel multi-talker dataset generated from LibriSpeech (LS)~\cite{panayotov2015librispeech}. Our experimental findings demonstrate that the UniX-Enc model, despite undergoing pretraining with a smaller and less diverse dataset, consistently outperforms the WavLM~\cite{chen2022wavlm} model combined with a BeamformIt~\cite{anguera2007beamformit} frontend. We will make our code and models publicly available upon publication.

\vspace{-.5em}
\section{Related Works}

\label{sec:related works}
\vspace{-.5em}
\subsection{Self-supervised Learning}
\vspace{-.5em}

Self-supervised learning represents a machine learning paradigm wherein the model derives its supervisory signal directly from the data itself. This approach has found extensive application in single-channel speech processing~\cite{oord2018representation,chung2019unsupervised,baevski2020wav2vec,hsu2021hubert,chen2022wavlm}. Typically, an SSL framework comprises two main stages: a pretraining phase, wherein the model learns speech representations through various pretext tasks, and a subsequent fine-tuning stage, where these acquired representations are optimized for specific downstream tasks. Within the realm of speech SSL models, our work is most closely related to WavLM~\cite{chen2022wavlm}. The WavLM model, an enhanced iteration of HuBERT~\cite{hsu2021hubert}, introduces several key modifications. First, it replaces convolutional relative position embedding with gated relative position bias. Second, it augments clean recordings with noise and interfering speech to enhance robustness. Third, it undergoes pretraining on larger and more diverse unlabeled speech datasets. In comparison to HuBERT, WavLM exhibits superior suitability for scenarios involving noise and speaker overlap.


\vspace{-.5em}
\section{UniX-Encoder}
\vspace{-.5em}
\label{sec:model}

Our proposed UniX-Encoder (UniX-Enc) adheres to the established self-supervised learning framework, encompassing two essential stages: pretraining and fine-tuning. In the pretraining phase, our model remains device-agnostic and task-agnostic, thus ensuring its adaptability to a broad spectrum of applications. Our model is pretrained on synthetic multi-channel multi-talker recordings and the mask prediction loss is chosen as the training objective.

In the subsequent fine-tuning phase, we leverage the UniX-Enc model as a versatile feature extractor for multi-channel recordings. This step enables us to enhance the acquired representations, tailoring them to specific downstream tasks, such as speech recognition and speaker diarization.


\vspace{-.5em}
\subsection{Pretraining stage}
\vspace{-.5em}

In the pretraining stage, we train the UniX-Enc model using synthetic multi-channel multi-talker speech data generated from LibriSpeech~\cite{panayotov2015librispeech}. The synthesis process is detailed in Algorithm \ref{algo:data simu}.

\begin{algorithm}[h]
\footnotesize
\caption{\footnotesize On-the-fly Multi-channel Multi-talker Speech Simulation}
\label{algo:data simu}
\begin{algorithmic}[1]
\STATE Given a batch of clean utterances $\mathcal{C} = \{\mathbf{u}^{i}\}_{i=1}^B$ with batch size $B$ and length $T$, a set of noises $\mathcal{N} = \{\mathbf{n}^{i}\}$, a set of room impulse responses (RIRs) $\mathcal{R} = \{\mathbf{RIR}^{i}\}$, the probability of mixing secondary speaker $p_i$ and mixing noise $p_n$, the range of microphone channels $[C_{min}, C_{max}]$, the range of length ratios for interferences $[L_{min}, L_{max}]$.
\STATE Select the number of microphone channels $C$ from $[C_{min}, C_{max}]$.
\FOR{each utterance $\mathbf{u}^{pri} \in \mathcal{C}$}
    \STATE Initialize three empty lists: $S$ for storing clean sources, $E$ for energy ratios, and $L$ for length ratios. $S.append(\mathbf{u}^{pri})$, $E.append(0)$, $L.append(1)$.
    \STATE Sample two values, $x$ and $y$, from a uniform distribution $\mathcal{U}(0,1)$.
    \IF{$x>p_i$} 
        \STATE Randomly select a secondary utterance $\mathbf{u}^{sec}$ ($\mathbf{u}^{sec} \neq \mathbf{u}^{pri}$) from $\mathcal{C}$, a mixing energy ratio $r_i$ from $\mathcal{U}(-6,6)$ and a length ratio $l_i$ from $\mathcal{U}(L_{min},L_{max})$. $S.append(\mathbf{u}^{sec})$, $E.append(r_i)$, $L.append(l_i)$.
    \ENDIF
    \IF{$y>p_n$}
        \STATE Randomly select a noise $\mathbf{n}$ from $\mathcal{N}$ and crop/repeat it to length L. Sample a mixing energy ratio $r_n$ from $\mathcal{U}(-5,20)$ and a length ratio $l_n$ from $\mathcal{U}(L_{min},L_{max})$. $S.append(\mathbf{n})$, $E.append(r_n)$, $L.append(l_n)$.
    \ENDIF
    \STATE Randomly select a RIR with $C$ channels from $\mathcal{R}$ and convolve it with the sources in $S$. The resulting list, containing the multi-channel reverberated sources, is denoted as $S^{r}$.
    \STATE Calculate the energy of multi-channel reverberated sources, denoting as $\{e_{i}^{r}\}_{i=1}^{len(S)}$. Rescale the sources in $S^{r}$ such that the energy of $i$th source $S^{r}[i]$ equals $-E[i] + e_{1}^{r} - e_{i}^{r}$. Subsequently, randomly crop the $S^{r}[i]$ to a length of $T \cdot L[i]$.
    \STATE Sum all the sources to create a multi-channel multi-talker mixture $\mathbf{m}$.
\ENDFOR
\end{algorithmic}
\end{algorithm}
\vspace{-1em}

\begin{figure}[t]
    \centering
    \includegraphics[width=0.8\linewidth]{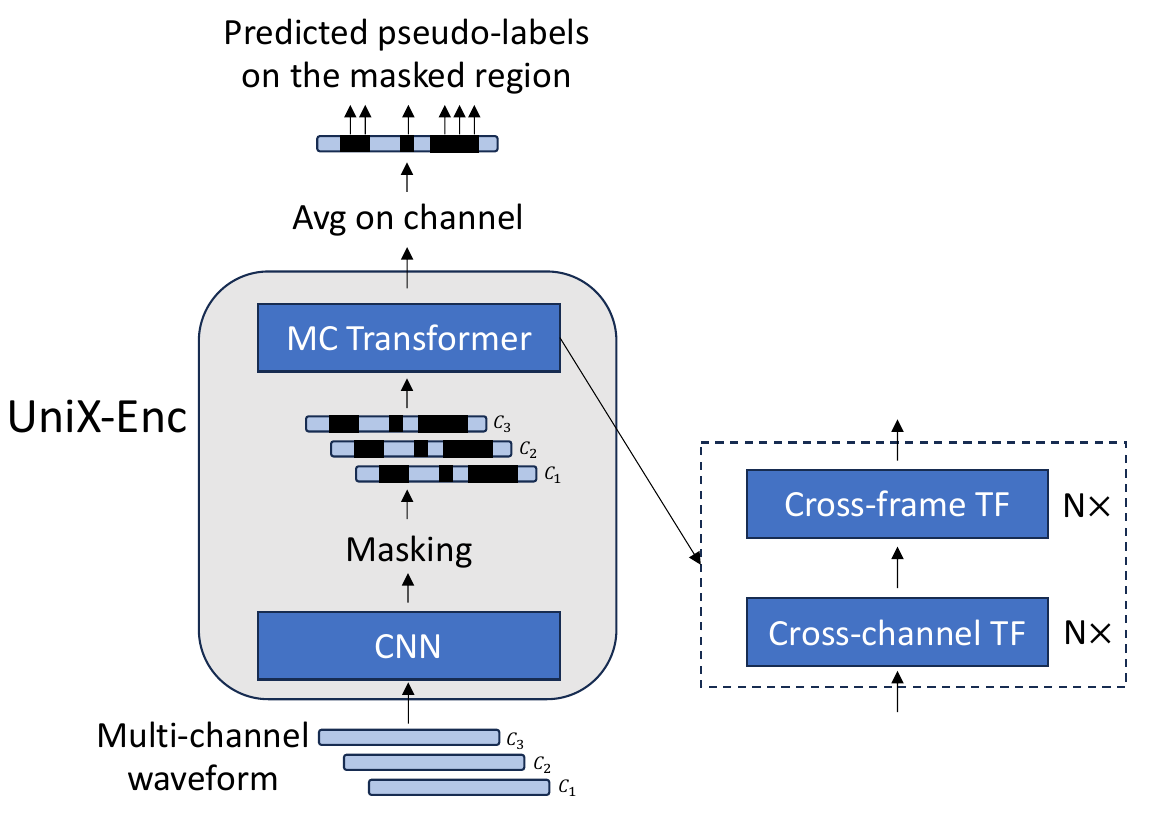}
    \vspace{-5mm}
    \caption{Model architecture of UniX-Encoder}
    \label{fig:model}
    \vspace{-5mm}
\end{figure}

The architecture of the UniX-Enc model is depicted in Fig.~\ref{fig:model}, and its network design draws inspiration from \cite{hsu2021hubert}. The UniX-Enc model takes a multi-channel waveform as its input, which is subsequently downsampled by a factor of 320 using a 7-layer CNN encoder. These downsampled representations undergo random masking in the time dimension, with the same time mask being applied across different channels.

Following this, the masked representations are passed through a multi-channel Transformer, comprising 6 cross-channel Transformers and 6 cross-frame Transformers. Each cross-channel Transformer consists of a cross-channel self-attention layer~\cite{wang2020neural, wang2021continuous, horiguchi2022multi} followed by a feed-forward network. The cross-channel self-attention layer calculates self-attention across the channel dimension, allowing the model to capture temporal differences between channels. To accommodate scenarios where the number of channels may be as minimal as 2, we incorporate a 1-frame context for both the key and value during the self-attention computation. The cross-frame Transformer operates similarly to the Transformer layer used in single-channel models, where we apply the same cross-frame attention and feed-forward mechanisms to process various microphone channels. Finally, the output of the multi-channel Transformer is averaged across the channel dimension, and a linear layer is employed to predict the pseudo-labels of the masked speech regions. 


For the training objective, we utilize the infoNCE loss~\cite{oord2018representation} as outlined in \cite{hsu2021hubert}. To better tailor the model for multi-talker applications, we incorporate the bi-label pretraining strategy described in \cite{huang2023self}, which involves predicting masked tokens for both the primary and secondary speakers. Our pretraining objective is formulated as
\begin{align}
   L_{pri}&=\sum_{t \in \mathcal{M}} -\log{\frac{\exp(\cos(\mathbf{o}_t^{pri} \cdot \mathbf{W}, \mathbf{e}_{k_t^{pri}})/\gamma)} {\sum_{k=1}^K \exp(\cos(\mathbf{o}_t^{pri} \cdot \mathbf{W}, \mathbf{e}_{k})/\gamma)}}\\
   L_{sec}&=\sum_{t \in \mathcal{M}} -\log{\frac{\exp(\cos(\mathbf{o}_t^{sec} \cdot \mathbf{W}, \mathbf{e}_{k_t^{sec}})/\gamma)} {\sum_{k=1}^K \exp(\cos(\mathbf{o}_t^{sec} \cdot \mathbf{W}, \mathbf{e}_{k})/\gamma)}}\\
   L &= L_{pri} + L_{sec}
\end{align}
where $\mathcal{M}$ is the set of masked timesteps, $\mathbf{o}_t^{pri}$ and $\mathbf{o}_t^{sec}$ are the deep representations of the primary and secondary speaker at timestep $t$, $\mathbf{W}$ is a projection matrix, $\mathbf{e}_k$ is the embedding corresponding to the pseudo-label $k\in\{1,...,K\}$, $k_t^{pri}$ and $k_t^{sec}$ are the ground truth pseudo-labels of the primary and secondary speakers at timestep $t$, $\text{cos}(\cdot,\cdot)$ computes the cosine similarity and $\gamma$ serves as a scaling factor for the logits. Notably, the pseudo-labels are derived through K-means clustering~\cite{macqueen1967some} applied to the 9th layer of the HuBERT Base model, with the total number of clusters $K$ set to 500.

\vspace{-.5em}
\subsection{Fine-tuning stage}
\vspace{-.5em}
\label{ssec:fine-tune}

\begin{figure}[t]
    \centering
    \includegraphics[width=0.8\linewidth]{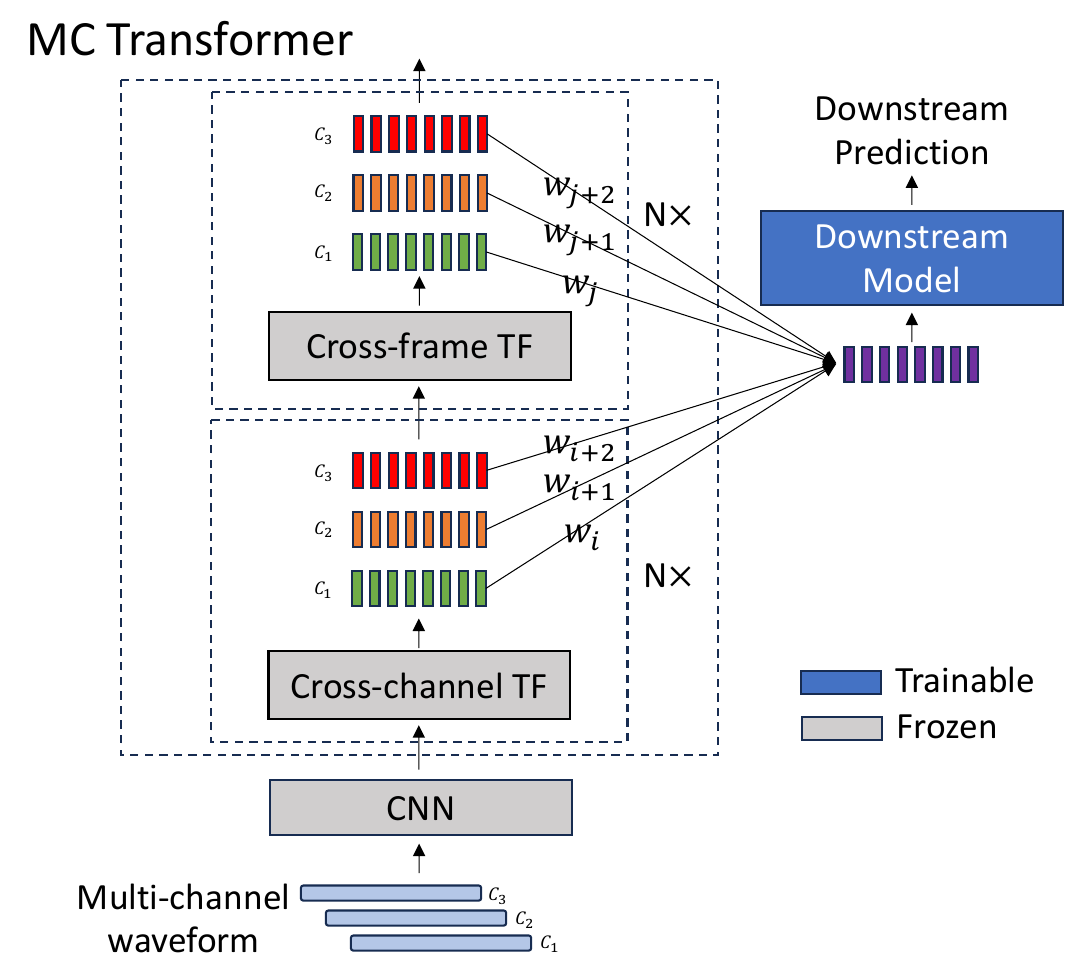}
    \vspace{-3mm}
    \caption{Fine-tuning stage of the UniX-Encoder}
    \label{fig:fine-tune}
    \vspace{-5mm}
\end{figure}

In the fine-tuning stage, various methodologies are available for leveraging the pretrained UniX-Enc model~\cite{baevski2020wav2vec,yang2021superb}. In this study, we adopted the ``weighted-sum" approach proposed in \cite{yang2021superb}. As illustrated in Fig. \ref{fig:fine-tune}, during fine-tuning, we forward the multi-channel speech through the pretrained UniX-Enc model and extract per-layer representations from the multi-channel Transformer. These representations are subsequently combined across different layers using a weighted average, as formulated by $\mathbf{F} = \sum_{i=1}^M w_i\cdot\mathbf{F}_i$, where $\mathbf{F}_i$ denotes the representation extracted from the $i$th layer and $w_i$ represents the corresponding learnable weight, such that $w_i \geq 0$ and $\sum_{i=1}^{M}w_i = 1$. The resulting representation $\mathbf{F}$ is then employed in various downstream models to fulfill diverse downstream tasks.

It is worth noting that in our UniX-Enc model, we conduct channel-wise averaging after the multi-channel Transformer. Consequently, the output of each cross-channel Transformer and cross-frame Transformer layer in the multi-channel Transformer has the shape $[C, T, D]$ rather than $[T, D]$, where $C$ is the number of channels, $T$ is the number of timesteps, and $D$ is the representation dimension. For each output from the cross-channel/cross-frame Transformer layers, we evaluate the effectiveness of two strategies: (1) averaging along the channel dimension and learning a single weight for it and (2) treating each channel as an individual layer and learning $C$ weights to combine them. Our experiments in Section \ref{sssec:ablation} reveal that both strategies yield comparable performance, and for the remainder of this paper, we opt for the latter strategy.

\vspace{-.5em}
\section{EXPERIMENTS}

\label{sec:exp}
\vspace{-.5em}
\subsection{Experimental settings}
\vspace{-.5em}
\label{ssec:exp_setting}
\subsubsection{Data}
\vspace{-.5em}
\label{sssec:subsubhead}
Our UniX-Enc model was pretrained on a synthetic multi-channel multi-talker dataset, with the simulation procedure detailed in Algorithm \ref{algo:data simu}. In this algorithm, the clean utterances are sourced from LibriSpeech (LS) 960h~\cite{panayotov2015librispeech}, while the noises are drawn from the DNS challenge dataset~\cite{reddy2021interspeech}. The number of microphone channels in our simulated dataset ranges from 2 to 4. We generated a total of $100,000$ room impulse responses (RIRs) for each possible number of microphone channels using the Pyroomacoustics library~\cite{scheibler2018pyroomacoustics}. The virtual room dimensions, including length, width, and height, span from 3 to 8 meters, 3 to 8 meters, and 2.5 to 4 meters, respectively. Since we do not assume any specific microphone topology during pretraining, we randomly sampled the positions of the microphones in the virtual space. The maximum distance between the geometric center of the microphone arrays and each microphone was constrained to fall within the range of 5cm to 15cm. Additionally, the room's reverberation time (RT60) was uniformly sampled between 0.05 and 0.8. Regarding the mixing process, we set the probability of mixing in a secondary speaker $p_i$ and mixing in noise sources $p_n$ to be 0.5. The range of length ratios for interferences is set to $[0.1, 0.5]$.

During the fine-tuning phase, we work with three distinct datasets: one for fine-tuning the pretrained model, another for validation and model selection, and the third for testing and performance comparison. These datasets are respectively referred to as the fine-tuning set, development set, and test set. They are all generated using the identical procedure outlined in Algorithm \ref{algo:fine-tuning}. The clean utterances for the fine-tuning, development, and test sets are drawn from LS train-clean-100, dev-clean, and test-clean, while the noise sources are derived from the train, dev, and test partitions of the WHAM! noise dataset~\cite{wichern2019wham}. The microphone configuration employed for our simulation aligns precisely with that described in \cite{chen2020continuous}, consisting of 6 microphones uniformly distributed around a circle and one positioned at the center. We synthesize 10 hours of audio for the fine-tuning set, and the number of utterances in the development and test sets matches that of LS dev-clean and test-clean. 



\begin{algorithm}[h]
\footnotesize
\caption{Fine-tuning/development/test set simulation}
\label{algo:fine-tuning}
\begin{algorithmic}[1]

\STATE Given a set of clean utterances $\mathcal{U} = \{\mathbf{u}^{i}\}$, a set of noises $\mathcal{N} = \{\mathbf{n}^{i}\}$, the range of RT60 $[R_{min}, R_{max}]$, the range of SIR $[I_{min}, I_{max}]$, the range of SNR $[N_{min}, N_{max}]$, the number of recordings to generate $M$.
\STATE $j \leftarrow 1$
\WHILE{$j \leq M$}
\STATE Randomly select a primary utterance $\mathbf{u}_1$ and a secondary utterance $\mathbf{u}_2$ from $\mathcal{U}$ and a noise $\mathbf{n}$ from $\mathcal{N}$.
\STATE Randomly sample a virtual room dimension and the RT60.
\STATE Randomly sample the positions of microphone array and sound sources in the virtual room. Compute the RIR and convolve it with $\mathbf{u}_1$, $\mathbf{u}_2$ and $\mathbf{n}$. The resulting sources are denoted as $\mathbf{u}_1^{r}$, $\mathbf{u}_2^{r}$ and $\mathbf{n}^{r}$.

\STATE Sample SIR $r_i$ from $\mathcal{U}(I_{min}, I_{max})$ and SNR $r_n$ from $\mathcal{U}(N_{min}, N_{max})$. Scale $\mathbf{u}_2^{r}$ and $\mathbf{n}^{r}$ according to $r_i$ and $r_n$.
\STATE Set the start time of $\mathbf{u}_1^{r}$ to 0. Randomly sample the starting time of $\mathbf{u}_2^{r}$ from $\mathcal{U}(0, \text{len}(\mathbf{u}_1^{r}))$. Repeat $\mathbf{n}^{r}$ to match the total length.  
\STATE Sum all the sources to create a multi-channel multi-talker mixture $\mathbf{m}$.
\STATE $j \leftarrow j + 1$
\ENDWHILE
\end{algorithmic}
\end{algorithm}
\vspace{-2em}
\subsubsection{Pretraining configuration}
\vspace{-.5em}
We pretrain our UniX-Enc model for 400,000 steps on the synthetic multi-channel multi-talker dataset. Within the initial 32,000 steps, the learning rate linearly increases to $5e^{-4}$, after which it linearly decreases to 0. Our model is trained on 16 NVIDIA Tesla V100 GPUs using the Adam optimizer~\cite{kingma2014adam}. We employ gradient accumulation for 2 steps. 

\vspace{-1em}
\subsubsection{Fine-tuning configuration}
\vspace{-.5em}



We evaluate our UniX-Enc model's performance in two distinct downstream tasks: speech recognition and speaker diarization. To facilitate this assessment, we generate similar multi-channel, multi-talker datasets for each task, with minor variations in hyperparameter settings. For speech recognition, to eliminate ambiguity regarding which speaker to transcribe, we enforce a requirement that the energy of the primary speaker significantly exceeds that of the secondary speaker. We set the range for RT60 between 0.1 and 0.6, while SIR and SNR vary from 5 to 20. In the context of speaker diarization, all speakers are considered equally important. For this task, the range for RT60 is set to $[0.05, 0.8]$, while SIR spans from -6 to 6, and SNR ranges from -5 to 20.

\textbf{Speech recognition}: In our speech recognition task, we employ a 2-layer Bidirectional Long Short-Term Memory (BLSTM) architecture with 1,024 hidden units as the downstream model. The output from the BLSTM layers is further processed through a linear layer to obtain the character distribution. We utilize the Connectionist temporal classification (CTC) loss~\cite{graves2006connectionist} as our training objective.

The fine-tuning of the downstream model is carried out on 4 NVIDIA RTX 3090 GPUs for 35,000 steps, using the Adam optimizer. We perform a learning rate sweep ranging from $1e^{-5}$ to $1e^{-2}$ and select the model that yields the best word error rate (WER) on the development set.

\textbf{Speaker diarization}: For our speaker diarization task, we employ a single-layer LSTM with 512 hidden units as the downstream model. The output of the LSTM layer is transformed into frame-level speaker activity using a linear layer. Our training objective involves binary cross-entropy loss with permutation-invariant training (PIT)~\cite{yu2017permutation, fujita2019end}.

The fine-tuning process for this model takes place on 4 NVIDIA RTX 3090 GPUs for 15,000 iterations, with the Adam optimizer. We perform a learning rate search across the range of $1e^{-5}$ to $1e^{-2}$ and select the model that attains the highest accuracy in frame-level speaker activity on the development set. 

\vspace{-1em}
\subsection{Results}
\vspace{-.5em}
\label{ssec:results}

\subsubsection{Speech recognition and speaker diarization}
\vspace{-.5em}
\begin{table}[t]
\footnotesize
\tabcolsep = 1.5mm
    \centering
    \caption{Speech recognition and speaker diarization performance on  multi-channel multi-talker LibriSpeech}
    \begin{tabular}{ccccccc} \toprule
        \multirow{2}{*}{Upstream} & Upstream & \multirow{2}{*}{Channel} & \multicolumn{2}{c}{WER (\%)} & \multicolumn{2}{c}{DER (\%)} \\
        & \#Params & & w/o LM & w. LM & dev & test \\
        \hdashline[1pt/2pt]\hdashline[0pt/1pt]
        WavLM Base & 94M & 0 & 27.55 & 22.28 & 19.54 & 19.47 \\
        WavLM Base+ & 94M & 0 & 24.07 & 19.07 & 16.54 & 17.08 \\
        WavLM Large & 315M & 0 & 14.32& 10.36 & 10.85	& 10.84\\
        \hdashline[1pt/2pt]\hdashline[0pt/1pt]
        WavLM Base & 94M & BF 1,0,4 & 25.78 & 20.48  & 21.10 & 21.25 \\
        WavLM Base+ & 94M & BF 1,0,4 & 21.81 &  17.11 & 14.94 & 15.60 \\
        WavLM Large & 315M & BF 1,0,4 & 13.43 & 9.66 & 11.16 & 11.46  \\
        \hdashline[1pt/2pt]\hdashline[0pt/1pt]
        UniX-Enc Base & 95M & 1,0,4 & 21.96 & 16.25 & 6.25 & 6.36 \\
        \bottomrule
    \end{tabular}
    \label{tab:asr_diar}
    \vspace{-5mm}
\end{table}

The performance of our model on the downstream tasks of speech recognition and speaker diarization is presented in Tables \ref{tab:asr_diar}. We selected three microphone channels (1, 0, and 4) from a total of seven, forming a linear microphone array configuration. As part of our baseline evaluation, we compare the performance of WavLM Base, Base+, and Large models in two scenarios: first, by extracting representations from a single channel (channel 0), and second, by performing beamforming with the BeamformIt toolkit~\cite{anguera2007beamformit} and extracting representations from the resulting beamformed audio.


It's important to note that comparing our UniX-Enc base model with WavLM Base+ and WavLM Large may not be entirely fair. This discrepancy arises from the fact that the latter models were pre-trained on a much larger dataset of 94k hours of diverse speech, while our UniX-Enc base and WavLM base models were only pre-trained on 960 hours of LibriSpeech \cite{panayotov2015librispeech}. Additionally, the WavLM Large model boasts a significantly larger model size.



In the context of speech recognition, we present WER results both with and without the inclusion of a language model (LM). Specifically, we employ the official LibriSpeech 4-gram LM. Our UniX-Enc Base model demonstrates remarkable performance, achieving a WER of 21.96\% without the LM and 16.25\% with the LM. These results not only surpass the performance of WavLM Base combined with the BeamformIt front-end but also rival the performance of WavLM Base+, a model pretrained on a significantly larger corpus of unlabeled data.

In the domain of speaker diarization, the performance of UniX-Enc stands out as extraordinary. It surpasses the WavLM Large model by a substantial margin, achieving a remarkable 4.5\% reduction in Diarization Error Rate (DER).

Two noteworthy observations emerge from our findings. Firstly, BeamformIt consistently enhances speech recognition performance, yet it occasionally has a detrimental impact on speaker diarization results. Secondly, the UniX-Enc Base model exhibits a particularly significant improvement in the speaker diarization task. We postulate that the reason behind the first observation (1) lies in the inherent differences between the two tasks. Unlike speech recognition, where a dominant speaker with higher energy and longer duration is present, speaker diarization lacks a clear "primary speaker." Consequently, in overlapping speech regions, BeamformIt may face challenges in determining which speaker to prioritize for enhancement. Regarding the second observation (2), our UniX-Enc model has been optimized specifically for multi-talker scenarios, guided by a bi-label pretraining objective. This optimization encourages the model to extract distinct speech representations from different sources, thus contributing to its impressive performance in speaker diarization tasks.


\vspace{-1em}
\subsubsection{Robustness to microphone topology}
\vspace{-.5em}


Table \ref{tab:channel} presents a comparative analysis of speaker diarization performance between our UniX-Enc and the combination of WavLM Base+ with a BeamformIt front-end across various microphone topologies. The results clearly indicate that our proposed UniX-Enc outperforms its counterpart in terms of DER. Remarkably, irrespective of the number and configuration of microphone channels, the UniX-Enc Base model consistently achieves a DER below 7\%. These findings underscore the model's impressive robustness across diverse microphone topologies.

\begin{table}[t]
    \centering
    \caption{DER (\%) for different microphone topologies}
    \begin{tabular}{c|c|c}
        Microphone & BeamformIt & \multirow{2}{*}{UniX-Enc Base} \\
        topology & WavLM Base+ & \\
        \hline
        1ch & 17.08 & 8.62 \\
        2ch & 16.07 & 6.36 \\
        3ch (linear) & 15.60 & 6.36 \\
        3ch (triangular) & 17.49 & 6.15 \\
        5ch & 16.17 & 6.72 \\
        \hline
    \end{tabular}
    \label{tab:channel}
    \vspace{-5mm}
\end{table}

\vspace{-1em}
\subsubsection{Ablation study}
\vspace{-.5em}
Furthermore, we conducted an ablation study to assess the influence of bi-label pretraining and same-layer weight averaging (the first strategy detailed in Section \ref{ssec:fine-tune}). As depicted in Table \ref{tab:ablation}, the incorporation of bi-label pretraining leads to a notable enhancement in speaker diarization performance, manifesting as a substantial reduction in the DER from 9.01\% to 6.36\%. Nevertheless, this advantageous effect comes with a slight trade-off, as it contributes to a marginal increase in the WER of approximately 1\% in speech recognition. These findings affirm the superior suitability of the bi-label pretraining objective for multi-talker scenarios. Notably, same-layer weight averaging exhibited minimal impact on both speech recognition and speaker diarization performance.

\label{sssec:ablation}
\begin{table}[t]
    \centering
    \caption{Impact of bi-label pretraining and same-layer weight average on the speech recognition and speaker diarization tasks}
    \begin{tabular}{c|cc|cc} 
        \multirow{2}{*}{System} & \multicolumn{2}{c}{DER} & \multicolumn{2}{|c}{WER}\\
        & DEV & TEST & w/o LM & w. LM \\
        \hline
        Full UniX-Enc model & 6.25 & 6.36 & 21.96 & 16.25 \\
        - bi-label & 8.78 &	9.01 & 20.78 & 15.48 \\
        + same-layer weight avg & 6.38 & 6.57 & 21.78 & 16.31 \\ 
        \hline
    \end{tabular}
    \label{tab:ablation}
    \vspace{-5mm}
\end{table}

\vspace{-1em}
\section{Conclusion and future work}
\vspace{-.5em}
\label{sec:conclusion}

In this paper, we introduce the \textbf{UniX-Encoder}, a self-supervised learning-based framework designed for ad-hoc microphone array speech processing. The UniX-Encoder serves as an efficient feature extractor for multi-channel speech data. We substantiate the effectiveness of our proposed model by conducting evaluations on two downstream tasks: speech recognition and speaker diarization. Our UniX-Encoder consistently outperforms the combination of WavLM and a BeamformIt frontend. In our future work, we plan to further evaluate the performance of our model on real-world multi-channel datasets, such as AMI and CHiME.



\vfill\pagebreak

\bibliographystyle{IEEEbib}
\bibliography{strings,refs}

\end{document}